\documentstyle[12pt]{article}
\newcommand{\be}{\begin{equation}}
\newcommand{\ee}{\end{equation}}
\newcommand{\ba}{\begin{array}}
\newcommand{\ea}{\end{array}}
\newcommand{\bea}{\begin{eqnarray}}
\newcommand{\eea}{\end{eqnarray}}
\def\cn{\hbox{cn}}
\def\dn{\hbox{dn}}
\def\sn{\hbox{sn}}

\def\cn{\hbox{cn}}
\def\dn{\hbox{dn}}
\def\sn{\hbox{sn}}

\def\cn{\hbox{cn}}
\def\dn{\hbox{dn}}
\def\sn{\hbox{sn}}

\begin{document}
\begin{titlepage}

\vspace{3cm}

\begin{center}
{\bf\large Joint Institute for Nuclear Research \\
Bogoliubov Laboratory of Theoretical Physics}
\end{center}

\vspace{1cm}

\begin{center}
{\Large{\bf Isotropic oscillator
in the space of constant\\
positive curvature. Interbasis expansions.}}

\vspace{1.0cm}

{\bf Ye.M.Hakobyan
\footnote{yera@thsun1.jinr.dubna.su},
G.S.Pogosyan
\footnote{pogosyan@thsun1.jinr.dubna.su},
A.N.Sissakian and S.I.Vinitsky}

\vspace{2cm}

\hspace{4cm}{\bf\sl This work is devoted to the memory of \\

\hspace{6cm}our dear friend I.V. Lutsenko}

\end{center}

\vspace{1cm}

\begin{abstract}
The Schr\"odinger equation is thoroughly analysed for the isotropic oscillator
in the three-dimensional space of constant positive curvature in the
spherical and cylindrical systems of coordinates. The expansion coefficients
between the spherical and cylindrical bases of the oscillator are calculated.
It is shown that the relevant coefficients are expressed through the
generalised hypergeometric functions $_4F_3$ of the unit argument or $6_j$ Racah
symbols extended over their indices to the region of real values. Limiting
transitions to a free motion and flat space are considered in detail.

Elliptic bases of the oscillator are constructed in the form of expansion
over the spherical and cylindrical bases. The corresponding expansion
coefficients are shown to obey the three-term recurrence relations.
\end{abstract}

\vspace{1.5cm}

\end{titlepage}

\section{Introduction}

Starting with the classical works by Schr\"odinger [1], Stivenson [3]
and Infeld [2] systems with accidental degeneracy in spaces of
constant curvature have attracted attention of many researchers in connection
with nontrivial realization of hidden symmetry in these problems and with
possible applications, especially to constructing many-particle wave functions
[4], nonrelativistic models of quark systems [5] and solutions of
the two-center problem [6].

Essential advances in the theory of systems with accidental degeneracy
have been made by Nishino [7], Higgs [8], Leemon [9] and [10, 11, 12, 13].
It has been shown that the complete degeneracy of the spectrum
of the Coulomb problem and harmonic oscillator on the three-dimensional
sphere in the orbital and azimuthal quantum numbers is caused by an
additional integral of motion: an analog of
Runge--Lenz's vector (for the Coulomb potential) and an analog of Demkov's
tensor (for the oscillator). However, in contrast with the flat space the
integrals of motion for the Coulomb problem and isotropic oscillator do not
form the Lie algebra as the relevant commutators are nonlinear. The latter
does not allow one to restore, respectively, the algebra or the group of
hidden symmetry. Later in the works [14,15] it has been shown
that as an algebra of hidden symmetry one can use quadratic algebras
of the general type, the so-called Racah algebras.
Systems with hidden symmetry for the harmonic
potential and those of Winternitz--Smorodinsky's type in the
three-dimensional space of constant curvature were also studied by using
the technique of path integrals in the papers by Barut, Inomata and Junker
[16,17], Grosche [18] and Grosche et al. [19,20,21].

A possible way of determining a group of hidden symmetry of
systems with accidental degeneracy is determination of the expansion
coefficients between different bases obtained after the separation of
variables in the Schr\"odinger equation. Such interbasis expansions have
first been considered for the "sphere--cylinder" transitions (isotropic
oscillator on the sphere), "sphere--parabola" transitions and those between
spherical and elliptical bases (for the Coulomb potential on the sphere and
hyperboloid) in [14,15,25]. It has been shown in [19] that like
for the Helmholtz equation [22] variables in the Schr\"odinger equation
for the potential of the isotropic oscillator on the three-dimensional sphere
are separated into all the six orthogonal systems of coordinates: spherical,
cylindrical, sphero-conical, two elliptic and ellipsoidal coordinate systems.

The aim of the present paper is the description of solutions of the
Schr\"odinger equation in the spherical, cylindrical and two elliptic systems
of coordinates and the calculation of expansion coefficients between the
corresponding bases. Note that the solution of the Schr\"odinger equation
for the isotropic oscillator in the spherical system of coordinates was found
in [8,9,14] and in the cylindrical and elliptic systems of coordinates
are presented for the first time.

The paper is organised as follows. Section 2 presents some known
results related to the Schr\"odinger equation for the three-dimensional
space of constant curvature. Section 3 is devoted to the solution
of the Schr\"odinger equation for the potential of the isotropic oscillator
in the spherical and cylindrical systems of coordinates. Section 4 is the
calculation of coefficients of the interbasis expansion between spherical
and cylindrical bases using the explicit expression for the wave functions of
the isotropic oscillator. In Section 5 the elliptic bases of the isotropic
oscillator are constructed as expansion over the spherical and cylindrical
ones.

\section{The Schr\"odinger equation and integrals of motion}

The Schr\"odinger equation in the space of constant curvature has the form
\begin{eqnarray}
\left[- \frac{\hbar^2}{2\mu} \Delta_{LB} + V({\rm{\bf x}})\right]
\Psi = E \Psi.
\end{eqnarray}
where $\Delta_{LB}$ is the Laplace--Beltrami operator that in an
arbitrary system of coordinates is given by
\begin{eqnarray}
\Delta_{LB} = \frac{1}{\sqrt g} \frac{\partial}{\partial x^{i}}
{\sqrt g} g^{ik} \frac{\partial}{\partial x^{k}},
\,\,\,\,\,\,\, d s^2 = g_{ik} d x^i d x^k
\\[2mm]
g^{ik} = (g_{ik})^{-1},
\,\,\,\,\, g = det(g_{ik}) \quad (i,k=1,2,3).
\nonumber
\end{eqnarray}
Choosing a metric of the space of constant curvature in the form
($r^2 = x_i x_i$)
\begin{eqnarray}
g_{ik} =
\frac{1}{1+ {r^2}/{R^2}}
\left[\left(\delta_{ik} - \frac{x_i x_k}{r^2}\right)
+ \frac{1}{1+ {r^2}/{R^2}} \frac{x_i x_k}{r^2}\right],
\end{eqnarray}
we derive the following expression for the Laplace--Beltrami operator:
\begin{eqnarray}
\Delta_{LB}
=\left(1 + \frac{r^2}{R^2}\right)
\left[\left(\delta_{ik} + \frac{x_i x_k}{R^2}\right)
\frac{\partial^2}{\partial x_{i} \partial x_{k}}
+ \frac{x_i}{R^2}\frac{\partial}{\partial x_{i}}
\right]
=
- \left(P_i P_i + \frac{1}{R^2}L_{i} L_{i} \right)
\end{eqnarray}
where $R$ is the curvature radius,
\begin{eqnarray}
P_i
= - i \left[\frac{\partial}{\partial x_{i}}
+ \frac{x_i x_k}{R^2}\frac{\partial}{\partial x_{k}}
\right],
\,\,\,\,\,\,\,\,\,
L_{i}
= - i \epsilon_{ijk} x_j \frac{\partial}{\partial x_k}
= \epsilon_{ijk} x_i P_j
\end{eqnarray}
and the following commutation relations hold:
\begin{eqnarray}
[P_i, x_l] = - i\left(\delta_{il}+\frac{x_i x_l}{R^2}\right),
\,\,\,\,\,\,\,\,
[L_i, x_l] = - i\epsilon_{ijl} x_j
\end{eqnarray}

It is easily seen that within the limits of the flat space, i.e. as
$R\rightarrow\infty$, the operator $P_i$ corresponds to an ordinary
momentum operator, and the Laplace--Beltrami operator (2) turns into
an ordinary Laplace operator in the flat three-dimensional space
$E_3$.

The issue of generalising the problem of isotropic oscillator for
the spaces of constant curvature with the use of the conformally flat metric
in the classical mechanics has obviously been solved for the first time in
\cite{NI}, where, in particular, an additional integral of motion
characteristic of an oscillator interaction was found.
Later, in \cite{HI} it has been shown that if the metric of the curved
space is chosen in the form of (3), the role of the potential of isotropic
oscillator in the flat space is played by
\begin{eqnarray}
V(r) = \frac{\mu \omega^2 r^2}{2},
\end{eqnarray}
and an additional integral of motion has the form:
\begin{eqnarray}
D_{ik}   = \frac{1}{2}\left(
P_i P_k + P_k P_i \right) +
\frac{\mu^2\omega^2}{\hbar^2} x_i x_k,
\end{eqnarray}
which in the limit of large $R$ exactly transforms into Demkov's tensor
\cite{DEM}. For the operators $L_i$ and $D_{ij}$ the following
commutation relations are valid:
\begin{eqnarray*}
[D_{ij}, L_k] &=&  i  \left(\epsilon_{ikl} D_{jl}
+\epsilon_{jkm} D_{im}
\right),
\\[2mm]
[D_{ik}, D_{jl}] &=& \frac{i \mu^2 w^2}{\hbar^2}
\left(\delta_{li}L_{kj}+\delta_{lk}L_{ij}+\delta_{ij}L_{kl}
+\delta_{jk}L_{il}\right)+
\frac{i}{2 R^2}\bigg(\{L_{ij},D_{lk}\}
\\[2mm]
&+&\{L_{il},D_{kj}\}+\{L_{kj},D_{il}\}
+\{L_{kl},D_{ij}\}\bigg), \quad L_{ik} = x_i P_k- x_k P_i.
\nonumber
\end{eqnarray*}
where $\{ , \}$ means the anticommutator of two operators.

The three-dimensional space of constant positive curvature can also
be realised geometrically on the three-dimensional sphere $S_{3}$ of the
radius $R$, imbedded into the four-dimensional Euclidean space, i.e on
the hypersurface
$$
q_0^2 + {q_i}{q_i} = {\bf R}^2,
$$
where the coordinates $q_i$ change in the region ${q_i}{q_i}\leq R^2$ and
to each value of $q_i$ correspond two points on the sphere.
Relation between the coordinates $x_i$ in the tangent space and $q_\mu$
($\mu=0,1,2,3$) is given by
$$
q_i = \frac{x_i}{\sqrt{1 + r^2/R^2}}, \,\,\,\,
q_0 = \frac{R}{\sqrt{1 + r^2/R^2}}
$$
obtained under the mapping from the center of the three-dimensional
hypersphere onto the plane tangent to the "North pole". Such a parametrization
of the space of constant curvature is often called in literature the
"geodesic parametrization" \cite{KAD} and in a one-to-one manner reflects
only the hemisphere (in this case the upper one) or the sphere with identified
diametrically opposite points.

In the coordinates $q_\mu$ we have
\begin{eqnarray}
P_i = - \frac{1}{R} N_i =\frac{i}{R}
\left(q_i \frac{\partial}{\partial q_0} -
q_0 \frac{\partial}{\partial q_i}\right),
\,\,\,\,\,\,\,
L_{i} = - i \epsilon_{ijk} q_j \frac{\partial}{\partial q_k}
\end{eqnarray}
and
\begin{eqnarray*}
\Delta_{LB}
= - \frac{1}{R^2}\left(N_i^2 + L_i^2 \right)
\end{eqnarray*}
where the operators $L_i$ ¨ $N_i$ are generators of the group O(4)
\begin{eqnarray*}
[L_i, L_j] = i \epsilon_{ijk} L_k, \,\,\,\,\,\,
[L_i, N_j] = i \epsilon_{ijk} N_k, \,\,\,\,\,\,
[N_i, N_j] = i \epsilon_{ijk} L_k
\end{eqnarray*}
The potential of the isotropic oscillator is given by the symmetric function
\begin{eqnarray}
V(r) \equiv V(q) = \frac{\mu \omega^2}{2} \frac{q^2}{1-q^2/R^2},
\quad q^2=q_1^2+q_2^2+q_3^2
\end{eqnarray}
with respect to the upper and lower hemispheres, equals zero at the poles of
the sphere and has a singularity at the equator. An additional integral of
motion is given by the expression

\begin{eqnarray}
D_{ik}   = \frac{1}{2R^2}\left(
N_i N_k + N_k N_i \right) +
\frac{\mu^2\omega^2}{\hbar^2} \frac{q_i q_k}{1-q^2/R^2}
\end{eqnarray}
and like in the case of the flat space leads to separation of variables
in the Schr\"odinger equation in more than one system of coordinates.

\section{Solution of the Schr\"odinger equation}

\subsection{Spherical basis}

In the spherical system of coordinates
\begin{eqnarray}
\begin{array}{ll}
q_1 = R \sin \chi \sin\vartheta \cos \varphi, &
q_2 = R \sin \chi \sin \vartheta \sin \varphi,
\nonumber\\
q_3 = R \sin \chi \cos \vartheta,  &
q_0 = R \cos \chi,
\end{array}
\end{eqnarray}
$$
0\leq
\chi \leq \pi \,\, , \, \,  0 \leq \vartheta \leq \pi
\,\, , \,\,  0 \leq \varphi < 2\pi,
$$
the oscillator potential has the form
\begin{eqnarray*}
V =
\frac{\mu\omega^2 R^2}{2} {\rm tg}^2\chi.
\end{eqnarray*}
Choosing the wave function according to
\begin{eqnarray}
\Psi(\chi,\vartheta,\varphi; R) = \frac{1}{\sqrt{R^3}}
Z(\chi)Y_{lm}(\vartheta,\varphi),
\,\,\,\,\,\,\,
l\in\mbox{{\rm\bf{N}}},
\,\,\,\,
m\in{\mbox{\rm\bf{Z}}},
\end{eqnarray}
where $Y_{lm}(\vartheta,\varphi)$ is an ordinary spherical function \cite{V},
after separation of variables in the Schr\"odinger equation we have
\begin{eqnarray}
\Biggl\{
\frac{1}{\sin^2\chi}
\frac{\partial}{\partial \chi}
\sin^2\chi \frac{\partial}{\partial \chi} +
\frac{2\mu R^{2}}{{\hbar}^2}
\Biggl[ E -
\frac{{\hbar}^2}{2\mu R^{2}}
\frac{l(l+1)}{\sin^{2}\chi} -
\frac{\mu\omega^2 R^2}{2} tg^{2}\chi\Biggr]\Biggr\}
Z(\chi;R) = 0
\end{eqnarray}
Then, introducing the notation
\begin{eqnarray}
\left(l + \frac{1}{2}\right)^2 = k_{1}^2, \,\,\,
\frac{\mu^2\omega^2 R^4}{{\hbar}^2} + \frac{1}{4}
= k_{2}^2, \,\,\,
\frac{2\mu R^2 E}{{\hbar}^2} + \frac{\mu^2 \omega^2 R^4}{{\hbar}^2}+ 1
= \cal{E}
\end{eqnarray}
and making the substitution
$$
Z(\chi) = \frac{f(\chi)}{\sin\chi},
$$
we arrive at the equation without the first derivative of the P\"oschl
--Teller-type
\begin{eqnarray*}
\frac{d^2 f}{d\chi^2} +
\Biggl[{\cal{E}} -
\frac{k_{1}^2-\frac{1}{4}}{\sin^{2}\chi} -
\frac{k_{2}^2-\frac{1}{4}}{\cos^{2}\chi}\Biggr]
f = 0,
\end{eqnarray*}
whose general solution is well known \cite{FL}.
The requirement of regularity of the wave function $Z(\chi)$ at $\chi=0$
and $\pi/2$ leads to quantization of the isotropic oscillator energy
\begin{eqnarray}
E^{\nu}_N (R) = \frac{\hbar^2}{2\mu}\Biggl[
\frac{(N+1)(N+3)}{R^2} + \frac{2\nu}{R^2}
\left(N + \frac{3}{2}\right)
\Biggr],
\end{eqnarray}
where the principal quantum number $N = 0,1,...$ is related with the radial
and orbital quantum numbers by $N = 2n_r + l$,
and the following notation is introduced:
$$
\nu \equiv k_2 - \frac{1}{2}
= \frac{1}{2}\sqrt{1+ \frac{4\mu^2 \omega^2 R^4}
{{\hbar}^2}} - \frac{1}{2}.
$$
Note that the degree of degeneracy, like in the case of motion in the
field of the harmonic isotropic oscillator in the three-dimensional
Euclidean space, is equal to $(N+1)(N+2)/2$.

The solution of the quasiradial Schr\"odinger equation (13), orthonormalised
in the interval $\chi \in [0, \frac{\pi}{2}]$ is
\begin{eqnarray}
Z(\chi) &\equiv& Z_{Nl}^{\nu}(\chi)
\nonumber\\[2mm]
&=&
\sqrt{\frac{2(N+\nu+2)(\frac{N-l}{2})!
\Gamma(\frac{N+l}{2}+\nu+2)}
{\Gamma(\frac{N+l+3}{2})
\Gamma(\frac{N-l+3}{2}+\nu)}}
(\sin\chi)^{l}(\cos\chi)^{\nu+1}
P_{\frac{N-l}{2}}^{(l + \frac{1}{2},
\nu + \frac{1}{2})} (\cos2\chi),
\end{eqnarray}
where $P_n^{(\alpha, \beta)}(x)$ are the Jacobi polynomials.

Let us consider the limit of the flat space. It is easily seen that at large
$R$ ($\nu\rightarrow \lambda R^2$, $\lambda = \mu\omega/\hbar^2$)
the formula (14) is used to restore the formula for the energy spectrum
of the three-dimensional oscillator
$$
\lim_{R\rightarrow\infty}
E^{\nu}_N (R) = E_N = \hbar\omega \left(N + \frac{3}{2}\right)
$$
Transition from the spherical system of coordinates on $S_3$ to the
relevant system of coordinates on $E_3$ is accomplished within the limit
$R\rightarrow\infty$, $\chi \rightarrow 0$ and $\chi \sim r/R$ where $r$
is the radius vector in the three-dimensional flat space ~\cite{IPSW}.
Using the known relation for the Jacobi polynomials \cite{BE}
\begin{eqnarray}
\lim_{\beta\rightarrow\infty}
P_{n}^{(\alpha,\beta)}\Bigl(1 - \frac{2x}{\beta}\Bigr)
=
L_{n}^{\alpha}(x),
\end{eqnarray}
where $L_{n}^{\alpha}(x)$ are the Laguerre polynomials, and taking account
of the limiting relations
$$
\sqrt{\frac{2(N+\nu+2)\Gamma(\frac{N+l}{2}+\nu+2)}
{\Gamma(\frac{N-l+3}{2}+\nu)}}
\,\, \frac{(\sin\chi)^{l}}{\sqrt{R^3}}
\Rightarrow
\sqrt{2\lambda^{3/2}}(\sqrt{\lambda} r)^l,\,\,\,
(\cos\chi)^{\nu+1/2}
\Rightarrow
e^{-\frac{\lambda r^2}{2}},\,\,\,
$$
we immediately get that
\begin{eqnarray}
\lim_{R\rightarrow\infty}
\, \frac{1}{\sqrt{R^3}} Z_{Nl}^{\nu}(\chi) =
\Biggl(\frac{\lambda}{\pi}\Biggr)^{1/4}
\sqrt{\frac{2^{l+1} \lambda (N-l)!!}{(N+l+1)!!}}
\,(\sqrt{\lambda} r)^{l}
\,e^{-\frac{\lambda r^2}{2}}
L_{\frac{N-l}{2}}^{l+\frac{1}{2}}(\lambda r^2) = R_{Nl}(r),
\end{eqnarray}
where $R_{Nl}(r)$ is the orthonormalised spherical radial wave function
of an ordinary three-dimensional isotropic oscillator in flat space \cite{FL}.

The second interesting limit is the transition to a free motion. As
$\nu\rightarrow 0$ ($\omega\rightarrow 0$) we have
$$
\lim_{\nu \rightarrow 0}
E_{N}^\nu (R) = E_{N}^0 (R) = \frac{\hbar^2}{2\mu}
\frac{(N+1)(N+3)}{R^2}
$$
Comparing the above-derived expression with the formula for the energy of a
free motion of particles on the sphere $\frac{\hbar^2 J(J+2)}{2\mu R^2}$
we get that $J=N+1$ and
$J=1,2,...$, and consequently, the ground state with
$J=0$ is missing in the limiting spectrum. As for the oscillator spectrum
$(N-l)$ is always even, within the limit of a free motion $(J-l)$
takes odd values and at fixed  $J$ there exist only states
with $l = J-1, J-3,....$ and, correspondingly, the degree of degeneracy of
the limiting spectrum is smaller than $(J+1)^2$, as it should be
for the free motion on the sphere.

Further, using the transformation \cite{S}
\begin{eqnarray}
x\,P_{m}^{(\lambda- \frac{1}{2}, \frac{1}{2})}(2x^2-1)
=
\frac{\Gamma(\lambda)\Gamma(m+3/2)}
{\Gamma(\lambda+m+1)\Gamma(1/2)}\,
C_{2m+1}^{\lambda}(x),
\end{eqnarray}
connecting odd Gegenbauer polynomials with the Jacobi polynomials
and passing from the quantum number $N$ to $J$, we come to the function
\begin{eqnarray*}
\lim_{\nu\rightarrow 0}
Z_{Nl}^{\nu}(\chi) =
\frac{2^{l+1}l!}{\sqrt{\pi}}
\sqrt{\frac{(J+1)(J-l)!}{(J+l+1)!}}
(\sin\chi)^{l}C_{J-l}^{l+1}(\cos\chi),
\end{eqnarray*}
which, with an accuracy to a factor, $\sqrt{2}$ corresponds to the solution
of the free Schr\"odinger equation on the three-dimensional sphere with the
impenetrable barrier at the equator ($\chi = \pi/2$).

\subsection{Cylindrical basis}

In the cylindrical system of coordinates
\begin{eqnarray}
\begin{array}{ll}
q_1 = R \sin \alpha \cos \phi_1,  &
q_2 = R \sin \alpha \sin \phi_1 ,\nonumber\\
q_3 = R \cos \alpha \sin \phi_2 , &
q_0 = R \cos \alpha \cos \phi_2 ,
\end{array}
\end{eqnarray}
$$
0\leq \alpha \leq \pi/2 , \, \, 0\leq \phi_1 < 2\pi, \,\,
- \pi \leq \phi_2 \leq \pi,
$$
the potential of the isotropic oscillator is written as
\begin{eqnarray}
V =
\frac{\mu\omega^2 R^2}{2}\left[\frac{1}{\cos^{2}\alpha
\cos^{2}\phi_2} - 1\right].
\end{eqnarray}
Choosing the wave function in the form
$$
\Psi(\phi_{1}, \alpha , \phi_{2}; R) = \frac{1}{\sqrt{R^3}}\,
\Phi(\alpha)\, K(\phi_{2})\,
\frac{e^{im \phi_{1}}}{\sqrt{2\pi}},
$$
after separation of variables we arrive at two differential equations of
the P\"oschl--Teller-type
\begin{eqnarray}
\frac{d^{2} M}{d{\alpha}^2} +
\left({\cal{E}} -
\frac{m^2 - \frac{1}{4}}{\sin^2{\alpha}}
- \frac{A - \frac{1}{4}}{\cos^2{\alpha}}
\right) M = 0,
\end{eqnarray}
\begin{eqnarray}
\frac{d^{2}K}{d{\phi}^2} +
\left(4A - \frac{k_2^2-\frac{1}{4}}{\sin^2{\phi}}
+ \frac{k_2^2-\frac{1}{4}}{\cos^2{\phi}}
\right) K = 0,
\end{eqnarray}
where $(\sin\alpha\cos\alpha)^{-1/2} M(\alpha) = \Phi(\alpha)$ ¨
$\phi = \frac{\phi_2}{2} + \frac{\pi}{4}$,
$\phi\in [0,\frac{\pi}{2}]$   ${\cal{E}}$ is determined by expression (14).
The spectrum of constants is determined by:
$$
A = (n_3 + \nu + 1)^2, \,\,\,\,\,\,\,
{\cal{E}} = (2n + n_3 + |m| + \nu + 2)^2,
$$
where the quantum numbers $n_3$ and $n$ run the values $0,1,2,...$.
Assuming the principal quantum number $N$ to be equal to $N = 2n + |m| + n_3$,
we get the formula (15) for the isotropic oscillator energy.
For the cylindrical basis we get the following expression:
\begin{eqnarray}
\Psi(\phi_{1}, \alpha, \phi_{2}; R) \equiv
\Psi_{Nmn_3}^{\nu}(\phi_{1}, \alpha, \phi_{2}; R) =
\frac{1}{\sqrt{R^3}}\Phi_{N|m|n_3}^{\nu}(\alpha)
K_{n_3}^{\nu}(\phi_{2})\frac{e^{im \phi_{1}}}{\sqrt{2\pi}}
\end{eqnarray}
where the functions $K_{n_{3}}^{\nu}(\phi_2)$ and $\Phi_{N|m|n_{3}}^{\nu}(\alpha)$,
normalised in the interval $\phi_2 \in [-\frac{\pi}{2}, \frac{\pi}{2}]$,
$\alpha\in [0,\frac{\pi}{2}]$, are
\begin{eqnarray}
\Phi_{N|m|n_{3}}^{\nu}(\alpha)
&=&
\sqrt{\frac{2(N+\nu+2)(\frac{N-|m|-n_3}{2})!
\Gamma(\frac{N+|m|+n_3}{2}+\nu+2)}
{(\frac{N+|m|-n_3}{2})!
\Gamma(\frac{N-|m|+n_3}{2}+\nu +2)}}
\nonumber\\
[2mm]
&\times&(\sin\alpha)^{|m|}(\cos\alpha)^{n_3+\nu+1}
P_{\frac{N-|m|-n_3}{2}}^{(|m|,\, n_3+\nu+1)}(\cos2\alpha),
\nonumber\\
[2mm]
K_{n_{3}}^{\nu}(\phi_2)
&=&
\frac{\sqrt{(n_3+\nu+1) \Gamma(n_3 + 2\nu + 2)(n_3)!}}
{2^{\nu+\frac{1}{2}} \Gamma(n_3 + \nu + \frac{3}{2})}
(\cos\phi_2)^{\nu+1}
P_{n_3}^{(\nu+ \frac{1}{2}, \nu + \frac{1}{2})}(\sin\phi_2).
\end{eqnarray}
Note that to a cylindrical system of coordinates corresponds an
additional integral of motion
\begin{eqnarray}
M = -  \frac{d^2}{d{\phi_2}^2}
+ \,\frac{\mu^2\omega^2 R^4}{\hbar^2}
\frac{1}{\cos^2{\phi_2}}
=
R^2 D_{33} + \frac{\mu^2\omega^2 R^4}{\hbar^2}
\end{eqnarray}

Within large $R$ the cylindrical system of coordinates on the sphere turns into
an ordinary cylindrical system of coordinates $(\rho, \phi, z)$ in the
Euclidean space $E_3$ \cite{IPSW}. Passing to the limit $R\rightarrow\infty$
and $\alpha, \phi_2 \rightarrow 0$, and assuming
\begin{eqnarray*}
\sin \alpha \sim \alpha \sim \frac{\rho}{R},\,\,\,\,\,\,\,
\sin \phi_2 \sim \phi_2 \sim \frac{z}{R},
\end{eqnarray*}
as well as using formula (17) and the following relation \cite{BE}
\begin{eqnarray*}
\lim_{\lambda\rightarrow\infty}
{\lambda}^{-n/2}C_{n}^{\lambda/2}
\left( t\sqrt{\frac{2}{\lambda}} \right)
= \frac{2^{-n/2}}{n!} {\cal{H}}_{n}(t),
\end{eqnarray*}
where ${\cal{H}}_{n}(z)$ are the Hermite polynomials  \cite{BE},
we get
\begin{eqnarray*}
\lim_{R\rightarrow\infty}
\frac{1}{R} \Phi_{N|m|n_{3}}^{\nu}(\alpha)
&=& \sqrt{\frac{2\lambda (\frac{N-|m|-n_3}{2})!}
{(\frac{N+|m|-n_3}{2})!}}
e^{-\frac{\lambda\rho^2}{2}}
(\sqrt{\lambda}\rho)^{|m|}
L_{\frac{N-|m|-n_3}{2}}^{|m|}(\lambda\rho^2),
\\
\lim_{R\rightarrow\infty}
\frac{1}{\sqrt{R}}
K_{n_{3}}^{\nu}(\phi_2)
&=&
\Biggl(\frac{\lambda}{\pi}\Biggr)^{1/4}
\frac{e^{-\frac{\lambda z^2}{2}}}
{\sqrt{2^{n_3}(n_3)!}}
{\cal{H}}_{n_3}(\sqrt{\lambda} z).
\end{eqnarray*}
Thus, formula (23) leads to the orthonormalised cylindrical basis
of the isotropic oscillator in the flat space.

Within the limit of a free motion $\nu\rightarrow 0$
$(\omega\rightarrow 0)$ with the use of the formula [28]
\begin{eqnarray*}
P_{n}^{(\frac{1}{2}, \, \frac{1}{2})}(x)
=
\frac{(2n+1)!!}{2^n (n+1)!}
\,
\frac{\sin[(n+1) \arccos x]}{\sin (\arccos x)},
\end{eqnarray*}
we have
\begin{eqnarray*}
\lim_{\nu \rightarrow 0}
K_{n_{3}}^{\nu}(\phi_2)
&=&
\sqrt{\frac{2}{\pi}}\,
\sin\{(n_3 + 1)(\phi_2 + \pi/2)\},
\\[2mm]
\lim_{\nu \rightarrow 0}
\Phi_{N|m|n_{3}}^{\nu}(\alpha)
&=&
\sqrt{\frac{(N+2)(\frac{N-n_3-|m|}{2})!
(\frac{N+n_3+|m|}{2}+1)!}
{(\frac{N-|m|+n_3}{2}+1)!
(\frac{N+|m|-n_3}{2})!}}
\\[2mm]
&\cdot&
(\sin\alpha)^{|m|}(\cos\alpha)^{n_3+1}
P_{\frac{N-n_3-|m|}{2}}^{(|m|,\, n_3+1)}(\cos2\alpha).
\end{eqnarray*}
Assuming $n_3 + 1 = |m_2|$, $|m| = |m_1|$ and $J=N+1$, we obtain
(with an accuracy to a factor $\sqrt{2}$) odd solutions of the Schr\"odinger
equation for a free motion in the cylindrical system of coordinates.

\section{Expansion between spherical and cylindrical bases}

\subsection{Calculation of the transition coefficients}

Let us write the expansion between the spherical and cylindrical bases
of the isotropic oscillator in the form
\begin{eqnarray}
\Psi_{Nlm}^{\nu}(\chi,\vartheta,\varphi; R)
=
\sum_{n_3= 0,1}^{N-|m|}
W_{Nlm}^{n_3}(\nu)
\Psi_{Nmn_3}^{\nu}(\phi_{1}, \alpha , \phi_{2}; R),
\end{eqnarray}
where the quantum number $n_3$ takes even and odd values depending
on parity $N-|m|$.

To calculate an explicit form of the expansion coefficients
$W_{Nlm}^{n_3}(\nu)$  it is sufficient to use orthogonality in
one of the variables for the functions entering into the cylindrical
wave function and to fix at the most appropriate point the second
variable that does not participate in integration.
Passing beforehand in the left-hand side of the expansion (26)
from the spherical coordinates to the cylindrical ones,
according to the formulae
\begin{eqnarray*}
\cos\chi  = \cos\alpha\cdot\cos\phi_{2}, \,\,\,\,\,\,
\sin\vartheta = \frac{\sin\alpha}
{\sqrt{1-\cos^{2}\alpha\cos^{2}\phi_{2}}}, \,\,\,\,\,\,
\phi =\phi_{1}
\end{eqnarray*}
and taking into account that as $\alpha\rightarrow 0$
\begin{eqnarray*}
\cos\chi \rightarrow\cos\phi_{2}, \,\,\,\,\,\,
\sin\vartheta \rightarrow
\frac{\sin\alpha}{\sin \phi_{2}}
\rightarrow 0,
\end{eqnarray*}
we derive
\begin{eqnarray*}
Z_{Nl}^{\nu}(\chi)
\rightarrow
Z_{Nl}^{\nu}(\phi_{2})
\end{eqnarray*}
\begin{eqnarray*}
Y_{lm}(\vartheta, \phi)
\rightarrow
\frac{(-1)^{\frac{m+|m|}{2}}}{2^{|m|}|m|!}
\sqrt{\frac{2l+1}{2}\frac{(l+|m|)!}{(l-|m|)!}}
\frac{(\sin\alpha)^{|m|}}{(\sin\phi_{2})^{|m|}}
\frac{e^{im \phi}}{\sqrt{2\pi}}
\end{eqnarray*}
\begin{eqnarray*}
\Phi_{N|m|n_{3}}^{\nu}(\alpha)
\rightarrow
\sqrt{\frac{2(N+\nu+2)(\frac{N+|m|-n_3}{2})!
\Gamma(\frac{N+|m|+n_3}{2}+\nu+2)}
{(\frac{N-|m|-n_3}{2})!
\Gamma(\frac{N-|m|+n_3}{2}+\nu+2)}}
\frac{(\sin\alpha)^{|m|}}{|m|!}.
\end{eqnarray*}
Then, substituting the asymptotic formulae derived into the interbasis
expansion (26), reducing $(\sin\alpha)^{|m|}$ and using the orthogonality
of the functions $K_{n_{3}}^{\nu}(\phi_2)$ in the interval
$-\frac{\pi}{2} \leq \phi_{2} \leq \frac{\pi}{2}$,
we arrive at the following integral representation for the coefficients
$W_{Nlm}^{}(\nu)$:
\begin{eqnarray}
W_{Nlmn_3}^{}(\nu)
&=& (-1)^{\frac{m+|m|}{2}}
\sqrt{\frac{(2l+1)(n_{3}+\nu+1)}
{2^{|m|+\nu+1}\Gamma(n_3 + \nu+ \frac{3}{2})}
\frac{(l+|m|)!(\frac{N-|m|-n_3}{2})!}
{(l-|m|)!(\frac{N+|m|-n_3}{2})!}
\frac{(n_3)!(\frac{N-l}{2})!}{\Gamma(\frac{N+l}{2}+\frac{3}{2})}}
\nonumber\\[2mm]
&\cdot& \sqrt{\frac
{\Gamma(\frac{N-|m|+n_3}{2}+\nu+2)
\Gamma(\frac{N+l}{2}+\nu+2)
\Gamma(n_3 + 2\nu+2)}
{\Gamma(\frac{N+|m|+n_3}{2}+\nu+2)
\Gamma(\frac{N-l}{2}+\nu+\frac{3}{2})
}}
\,\,
A_{N|m|n_3}^{l}(\nu),
\end{eqnarray}
where
\begin{eqnarray}
A_{N|m|n_3}^{l}(\nu) =
{\int_{-\pi/2}^{\pi/2}
(\sin\phi_2)^{l-|m|}
(\cos\phi_2)^{2\nu+2}
P_{\frac{N-l}{2}}^{(l+\frac{1}{2},\nu+\frac{1}{2})}(\cos 2\phi_2)
P_{n_3}^{(\nu+\frac{1}{2},\nu+\frac{1}{2})}(\sin \phi_2)
d\phi_2}.
\end{eqnarray}

A complete solution of the problem needs calculation of the integral in
formula (27). Let us consider separately the cases of even and odd quantum
number $n_3$. Separating the interval of integration (28) into two intervals
$(- \frac{\pi}{2}, 0)$ and $(0, \frac{\pi}{2})$, after the substitution
in the first integral $\phi_2 \rightarrow -\phi_2$ we see that
the value of the integral is just doubled due to parity $(l-|m|-n_3)$.
Then, using the well-known transformation for the Jacobi polynomials
\cite{S}
$$
P_{n_3}^{(\alpha, \alpha)}(x) = \cases{
\displaystyle
\frac{\Gamma (n_3 + \alpha + 1)(\frac{n_3}{2})!}
{\Gamma (\frac{n_3}{2} + \alpha + 1)(n_3)!}\,\,
P_{\frac{n_3}{2}}^{(\alpha, \, - \frac{1}{2})}
(2x^2 - 1) & for $n_3$ -- even, \cr \cr \cr
\displaystyle
\frac{\Gamma (n_3 + \alpha + 1)\left( \frac{n_3 -1}{2} \right) !}
{\Gamma \left( \frac{n_3+1}{2} + \alpha \right) (n_3)!}\,
x
\,P_{\frac{n_3-1}{2}}^{(\alpha, \, \frac{1}{2})}
(2x^2 - 1) & for $n_3$ -- odd, \cr }
$$
after the substitution $x = \cos^2\phi_2$, we come to the following
two table integrals for even and odd $n_3$
\begin{eqnarray*}
A_{N|m|n_3}^{l(+)} (\nu)
&=&
\frac{(-1)^{\frac{n_3}{2}}}{2^{\nu + \frac{l-|m|}{2}+1}}  \cdot
\frac{\Gamma (n_3 + \nu + \frac{3}{2})(\frac{n_3}{2})!}
{\Gamma (\frac{n_3+3}{2} + \nu)(n_3)!}  \times
\\[2mm]
&\times&
{\int_{-1}^{1}
(1-x)^{\frac{l-|m|-1}{2}}
(1+x)^{\nu+\frac{1}{2}}
P_{\frac{N-l}{2}}^{(l+\frac{1}{2}, \nu+\frac{1}{2})}(x)
P_{n_3}^{(-\frac{1}{2},\nu+\frac{1}{2})}(x)
dx},
\\[3mm]
A_{N|m|n_3}^{l(-)} (\nu)
&=&
\frac{(-1)^{\frac{n_3-1}{2}}}{2^{\nu + \frac{l-|m|+3}{2}}} \cdot
\frac{\Gamma (n_3 + \nu + \frac{3}{2}) \left( \frac{n_3-1}{2} \right) !}
{\Gamma \left( \frac{n_3+2}{2} + \nu \right)(n_3)!}
\\[2mm]
&\times&
{\int_{-1}^{1}
(1-x)^{\frac{l-|m|}{2}}
(1+x)^{\nu+\frac{1}{2}}
P_{\frac{N-l}{2}}^{(l+\frac{1}{2},\nu+\frac{1}{2})}(x)
P_{\frac{n_3-1}{2}}^{(\frac{1}{2},\nu+\frac{1}{2})}(x)
dx}.
\end{eqnarray*}
Using the formula for integration of the two Jacobi polynomials\cite{P}
\begin{eqnarray*}
{\int_{-1}^{1}
(1-x)^{\tau}
(1+x)^{\beta}
P_{n}^{(\alpha,\beta)}(x)
P_{m}^{(\rho,\beta)}(x)
dx} =
\frac{2^{\beta+\tau+1}
\Gamma(\alpha-\tau+n)\Gamma(\beta+n+l)}
{(m)!(n)!\Gamma(\rho+1)\Gamma(\alpha-\tau)
}
\\[2mm]
\frac{\Gamma(\rho+m+1)\Gamma(\tau+1)}
{\Gamma(\beta+\tau+n+2)}
\,
{_4F_3}\left\{\left.\matrix{
-m,\,\,\rho+\beta+m+1,\,\,
\tau+1,\,\,\tau-\alpha+1\cr
\rho+1,\,\,\beta+\tau+n+2,\,\,
\tau-\alpha-n+1\cr}\right| 1\right\},
\end{eqnarray*}
we immediately get that $A_{N|m|n_3}^{l(\pm)} (\nu)$
is expressed through the generalised hypergeometric function
$_4F_3$ of the unit argument
\begin{eqnarray*}
A_{N|m|n_3}^{l(+)}(\nu)
&=&
\frac{(-1)^{\frac{n_3}{2}}}{\sqrt{\pi}}
\frac{
\Gamma(n_3+\nu+\frac{3}{2})\Gamma(\frac{N+|m|}{2}+1)
\Gamma(\frac{N-l+3}{2}+\nu)}
{(n_3)!\Gamma(\frac{n_3+3}{2}+\nu)(\frac{N-l}{2})!
\Gamma(\frac{l+|m|}{2}+1)
}
\frac{\Gamma(\frac{l-|m|+1}{2})
\Gamma(\frac{n_3+1}{2})}
{\Gamma(\frac{N-|m|}{2}+\nu+2)}
\\[2mm]
&\cdot& \,
{_4F_3}\left\{\left.\matrix{
-\frac{n_3}{2},\,\,\frac{n_3}{2}+\nu+1,\,\,
\frac{l-|m|+1}{2},\,\,-\frac{l+|m|}{2}\cr
\frac{1}{2},\,\,
\frac{N-|m|}{2}+\nu+2,\,\,
-\frac{N+|m|}{2}
\cr}\right| 1\right\},
\end{eqnarray*}
and analogously
\begin{eqnarray*}
A_{N|m|n_3}^{l(-)} (\nu)
&=&
\frac{(-1)^{\frac{n_3-1}{2}}}{\sqrt{\pi}}
\frac{2
\Gamma(n_3+\nu+\frac{3}{2})\Gamma(\frac{N+|m|+1}{2})
\Gamma(\frac{N-l+3}{2}+\nu)}
{(n_3)!\Gamma(\frac{n_3+2}{2}+\nu)(\frac{N-l}{2})!
\Gamma(\frac{l+|m|+1}{2})
}
\frac{\Gamma(\frac{l-|m|}{2}+1)
\Gamma(\frac{n_3}{2}+1)}
{\Gamma(\frac{N-|m|+5}{2}+\nu)}
\\[2mm]
&\cdot&\,
{_4F_3}\left\{\left.\matrix{
-\frac{n_3-1}{2},\,\,\frac{n_3+3}{2}+\nu,\,\,
\frac{l-|m|}{2}+1,\,\,-\frac{l+|m|-1}{2}\cr
\frac{3}{2},\,\,
\frac{N-|m|+5}{2}+\nu,\,\,
-\frac{N+|m|-1}{2}
\cr}\right| 1\right\}.
\end{eqnarray*}
Taking account of the known symmetry property for the series
${_4F_3}(1)$ of the Saalsch\"utz-type \cite{BA}
\begin{eqnarray*}
{_4F_3}\left\{\left.\matrix{
-n,\,\,b,\,\,c,\,\,d\cr
e,\,\,f,\,\,g\cr}\right| 1\right\} =
{{(f-b)_{n}(g-b)_{n}}
\over{(f)_{n}(g)_{n}}}
{_4F_3}\left\{\left.\matrix{
-n,\,\,b,\,\,e-c,\,\,e-d\cr
e,\,\,b-f-n+1,\,\,b-g-n+1\cr}\right| 1\right\},
\end{eqnarray*}
$$
-n + b + c + d = 1 + e + f + g,
$$
one can easily be convinced that both the hypergeometric functions
${_4F_3}(1)$ entering into $A_{N|m|n_3}^{l(\pm)} (\nu)$
can be transformed to a unique form:
\begin{eqnarray*}
{_4F_3}\left\{\left.\matrix{
-\frac{n_3}{2},\,\,\frac{n_3}{2}+\nu+1,\,\,
\frac{l-|m|+1}{2},\,\,-\frac{l+|m|}{2}\cr
\frac{1}{2},\,\,
\frac{N-|m|}{2}+\nu+2,\,\,
-\frac{N+|m|}{2}
\cr}\right| 1\right\}
=
\frac{(-1)^{\frac{n_3}{2}}\Gamma(\frac{1}{2})
\Gamma(\frac{l-|m|}{2}+1)\Gamma(\frac{l+|m|}{2}+1)}
{\Gamma(\frac{n_3 +1}{2})
\Gamma(\frac{l+|m|-n_3}{2}+1)\Gamma(\frac{l-|m|-n_3}{2}+1)}
\\[2mm]
\frac{\Gamma(\frac{n_3}{2}+\nu+\frac{3}{2})
\Gamma(\frac{N-|m|}{2}+\nu+2) (\frac{N+|m|-n_3}{2})!}
{\Gamma(\frac{N+|m|}{2}+1)\Gamma(\frac{N-|m|+n_3}{2}+\nu+2)
\Gamma(\nu+1)}
\cdot\,
{_4F_3}\left\{\left.\matrix{
-\frac{n_3}{2},\,\,-\frac{n_3-1}{2},\,\,
-\frac{N-l}{2},\,\,\frac{N+l}{2}+\nu+2\cr
\nu+\frac{3}{2},\,\,
\frac{l+|m|-n_3}{2}+1,\,\,
\frac{l-|m|-n_3}{2}+1
\cr}\right| 1\right\},
\end{eqnarray*}
\begin{eqnarray*}
{_4F_3}\left\{\left.\matrix{
-\frac{n_3-1}{2},\,\,\frac{n_3+3}{2}+\nu,\,\,
\frac{l-|m|}{2}+1,\,\,-\frac{l+|m|-1}{2}\cr
\frac{3}{2},\,\,
\frac{N-|m|+5}{2}+\nu,\,\,
-\frac{N+|m|-1}{2}
\cr}\right| 1\right\}
=
\frac{(-1)^{\frac{n_3-1}{2}}\Gamma(\frac{3}{2})
\Gamma(\frac{l-|m|+1}{2})\Gamma(\frac{l+|m|+1}{2})}
{\Gamma(\frac{n_3}{2}+1)
\Gamma(\frac{l+|m|-n_3}{2}+1)\Gamma(\frac{l-|m|-n_3}{2}+1)}
\\[2mm]
\frac{\Gamma(\frac{n_3}{2}+\nu+\frac{3}{2})
\Gamma(\frac{N-|m|}{2}+\nu+\frac{5}{2})
(\frac{N+|m|-n_3}{2})!}
{\Gamma(\frac{N+|m|+1}{2})\Gamma(\frac{N-|m|+n_3}{2}+\nu+2)
\Gamma(\nu+1)}
\cdot\,
{_4F_3}\left\{\left.\matrix{
-\frac{n_3}{2},\,\,-\frac{n_3-1}{2},\,\,
-\frac{N-l}{2},\,\,\frac{N+l}{2}+\nu+2\cr
\nu+\frac{3}{2},\,\,
\frac{l+|m|-n_3}{2}+1,\,\,
\frac{l-|m|-n_3}{2}+1
\cr}\right| 1\right\}.
\end{eqnarray*}
After simple transformations we finally derive the sought formula
for the coefficients of the interbasis expansion
$W_{Nlm}^{n_3}(\nu)$
\begin{eqnarray}
W_{Nlmn_3}^{n_3}(\nu)
=
\frac{(-1)^{\frac{m+|m|}{2}}\sqrt{\pi}}{2^{l+\nu+1}}
\frac{\sqrt{(2l+1)(n_3+\nu+1)(l+|m|)!(l-|m|)!}}
{\Gamma(\frac{l-|m|-n_3}{2}+1)\Gamma(\frac{l+|m|-n_3}{2}+1)
\Gamma(\nu+\frac{3}{2})}
\nonumber\\[2mm]
\cdot \sqrt{\frac
{\Gamma(\frac{N+l}{2}+\nu+2)
\Gamma(\frac{N-l+3}{2}+\nu)
\Gamma(n_3 + 2\nu+2)(\frac{N-|m|-n_3}{2})!(\frac{N+|m|-n_3}{2})!}
{\Gamma(\frac{N+l}{2}+\frac{3}{2})
(\frac{N-l}{2})!\Gamma(\frac{N-|m|+n_3}{2}+\nu+2)
\Gamma(\frac{N+|m|+n_3}{2}+\nu+2)(n_3)!}}
\nonumber\\[2mm]
\cdot \,
{_4F_3}\left\{\left.\matrix{
-\frac{n_3}{2},\,\,-\frac{n_3-1}{2},\,\,
-\frac{N-l}{2},\,\,\frac{N+l}{2}+\nu+2\cr
\nu+\frac{3}{2},\,\,
\frac{l+|m|-n_3}{2}+1,\,\,
\frac{l-|m|-n_3}{2}+1
\cr}\right| 1\right\}.
\end{eqnarray}
Note that the expression we have derived for $W_{Nlm}^{n_3}(\nu)$
is independent of parity of the quantum number $n_3$.

\subsection{Connection with the Racah coefficients}

The interbasis expansion coefficients (29) can also be expressed
through $6j$, the symbols or Racah coefficients of the SU(2) group,
extended over their indices to the region of real values. Comparing
the expression for $W_{Nmn_3}^{l}(\nu)$ with the representation of the
Racah coefficients $W(abed;\ cf)$ through the hypergeometric functions
${_4F_3}(1)$ of the unit argument \cite{V}
\begin{eqnarray}
W(abed;\ cf)
=
\frac{\Delta(abc)\Delta(cde)\Delta(aef)\Delta(bdf)
}
{(a+b-c)!(d+e-c)!
(a-f+e)!(b-f+d)!(c-a-d+f)!
}
\nonumber\\[2mm]
\frac{(a+b+d+e+1)!}
{(c-b-e+f)!}
{_4F_3}\left\{\left.\matrix{
-a-b+c,\,\,-b-d+f,\,\,
-a-e+f,\,\,c-d-e\cr
-a-b-d-e-1,\,\,-a+c-d+f+1,\,\,
-b+c-e+f+1\cr}\right| 1\right\},
\end{eqnarray}
where $\Delta(abc)$ is
$$
\Delta(abc) = \sqrt{\frac{(a+b-c)!(a-b+c)!
(b+c-a)!}{(a+b+c+1)!}},
$$
and taking account of the symmetry property of the Racah coefficients
$$
W(abed;\ cf) = i (-1)^{a+b-c} W(ab\bar e \bar d;\ c \bar f),
$$
$$
\bar e = - e - 1, \,\,\, \bar d = - d - 1, \,\,\, \bar f = - f - 1,
$$
after simple calculations we get the required formula
\begin{eqnarray}
W_{Nlm}^{n_3}(\nu) =
(-1)^{\frac{N-l}{2}+\frac{m+|m|}{2}}
\sqrt{(l+1/2)(n_3+\nu+1)}
W(ab ed;\ cf),
\end{eqnarray}
\begin{center}
\begin{tabular}{lll}
$a = \frac{N+|m|}{4}$, \,\,
&
$b = \frac{N-|m|-1}{4}$, \,\,
&
$c = \frac{2l-1}{4}$,
\\[2mm]
$d = \frac{N-|m|}{4} + \frac{\nu}{2} + \frac{1}{4}$, \,\,
&
$e = \frac{N+|m|}{4} + \frac{\nu}{2}$, \,\,
&
$f = \frac{n_3}{2} + \frac{\nu}{2}$. \\
\end{tabular}
\end{center}
Then, using the relation of orthonormalization for the Racah
coefficients \cite{V}
\begin{eqnarray*}
\sum_{c}\sqrt{(2c+1)(2f+1)} W(ab ed;\ cf) W(ab ed;\ c f^{\prime})
= \delta_{f f^{\prime}},
\end{eqnarray*}
we can write the inverse expansion in the form
\begin{eqnarray*}
\Psi_{Nmn_3}^{\nu}(\phi_{1}, \alpha , \phi_{2}; R)
=
\sum_{l=|m|, |m|+1}^{N}
{\tilde W}_{Nmn_3}^{l}(\nu)
\Psi_{Nlm}^{\nu}(\chi,\vartheta,\varphi; R),
\end{eqnarray*}
where summation over $l$ starts with $|m|$ or $|m|+1$ depending
on the parity of the number $N-|m|$, and the coefficients
\begin{eqnarray*}
{\tilde W}_{Nmn_3}^{l}(\nu)
=
W_{Nlm}^{n_3}(\nu)
\end{eqnarray*}
can be expressed through the polynomials ${_4F_3}(1)$ with the use of the
representation (30).

\subsection{Limiting relations}

Consider limiting transitions to the flat space and free motion in the
expansion coefficients $W_{Nmn_3}^{l}(\nu)$.

{\it 4.3.1} As $R\rightarrow \infty$ the generalised hypergeometric function
${_4F_3}(1)$ transforms into ${_3F_2}(1)$ according to
\begin{eqnarray*}
{_4F_3}\left\{\left.\matrix{
-\frac{n_3}{2},\,\,-\frac{n_3-1}{2},\,\,
-\frac{N-l}{2},\,\,\frac{N+l}{2}+\nu+2\cr
\nu+1,\,\,
\frac{l+|m|-n_3}{2}+1,\,\,
\frac{l-|m|-n_3}{2}+1
\cr}\right| 1\right\}
\Rightarrow
{_3F_2}\left\{\left.\matrix{
-\frac{n_3}{2},\,\,-\frac{n_3-1}{2},\,\,
-\frac{N-l}{2}\cr
\frac{l+|m|-n_3}{2}+1,\,\,
\frac{l-|m|-n_3}{2}+1
\cr}\right| 1\right\}
\end{eqnarray*}
Having made the relevant limiting transition in gamma functions, after
simple algebraic transformations we get the known formula for the
coefficients of the interbasis expansion between the spherical and cylindrical
bases of the harmonic isotropic oscillator in the flat Euclidean space
\cite{PT}:
\begin{eqnarray*}
\lim_{\nu \rightarrow\infty}W_{Nlm}^{n_3}(\nu)
=
\frac{(-1)^{\frac{m+|m|}{2}}}{2^{l-n_3}}
\sqrt{\frac{(N-|m|-n_3)!!(N+|m|-n_3)!!}
{(N+l+1)!!(N-l)!!(n_3)!}}
\\
\frac{\sqrt{(2l+1)(l+|m|)!(l-|m|)!}}
{\Gamma(\frac{l-|m|-n_3}{2}+1)\Gamma(\frac{l+|m|-n_3}{2}+1)}
\,
{_3F_2}\left\{\left.\matrix{
-\frac{n_3}{2},\,\,-\frac{n_3-1}{2},\,\,
-\frac{N-l}{2}\cr
\frac{l+|m|-n_3}{2}+1,\,\,
\frac{l-|m|-n_3}{2}+1
\cr}\right| 1\right\}.
\end{eqnarray*}
On the other hand, the limiting transition to the flat space can directly be
traced in the formula (31). Indeed, using at large
$R$ the asymptotic coupling \cite{V}
\begin{eqnarray*}
W(a b e+R, d+R; c, f+R) \approx
\frac{1}{\sqrt{2R(2c+1)}}
\, C_{a, \, f - e; \,b, \, d - f}^{c, \, d - e}
\end{eqnarray*}
and the symmetry property of the Clebsch--Gordan coefficients \cite{V}
\begin{eqnarray*}
C_{a, -\alpha; b, -\beta}^{c, - \gamma}
= (-1)^{a+b-c} \, C_{a, \alpha; \, b, \beta}^{c, \gamma},
\end{eqnarray*}
we obtain that
\begin{eqnarray*}
\lim_{\nu \rightarrow\infty}W_{Nlm}^{n_3}(\nu)
=
(-1)^{\frac{m+|m|}{2}}
\,
C_{\frac{N+|m|}{4}, \frac{N+|m|-2n_3}{4};
\frac{N-|m|-1}{4}, \frac{2n_3-N+|m|-1}{4}}
^{\frac{2l-1}{4}, \frac{2|m|-1}{4}}.
\end{eqnarray*}

{\it 4.3.2.} Assuming $\nu = 0$, ($\omega  = 0$) and passing to
the quantum numbers corresponding to the free motion on the sphere
$m=m_1, n_3+1=|m_2|, N=J-1$ (note that $J-l$
is odd and  $J-|m_1|-|m_2|$ is even), we have
\begin{eqnarray}
\lim_{\nu\rightarrow 0} W_{Nlm}^{n_3}(\nu)
= (-1)^{\frac{m_1 + |m_1|}{2}}
\frac{|m_2|}{2^l}
\sqrt{\frac
{(\frac{J+|m_1|-|m_2|}{2})!(\frac{J-|m_1|-|m_2|}{2})!
(J+l+1)!!(J-l)!!}
{(\frac{J+|m_1|+|m_2|}{2})!(\frac{J-|m_1|+|m_2|}{2})!
(J+l)!!(J-l-1)!!}}
\nonumber\\[3mm]
\frac{\sqrt{(l+1/2)(l+|m|)!(l-|m|)!}}
{\Gamma(\frac{l-|m_1|-|m_2|}{2}+\frac{3}{2})
\Gamma(\frac{l+|m_1|-|m_2|}{2}+\frac{3}{2})}
\cdot \,
{_4F_3}\left\{\left.\matrix{
-\frac{|m_2|-1}{2},\,\,-\frac{|m_2|-2}{2},\,\,
-\frac{J-l-1}{2},\,\,\frac{J+l+3}{2}\cr
\frac{3}{2},\,\,
\frac{l+|m|-|m_2|}{2}+\frac{3}{2},\,\,
\frac{l-|m|-|m_2|}{2}+\frac{3}{2}
\cr}\right| 1\right\}
\nonumber\\[3mm]
= (-1)^{\frac{J-l-1}{2}+\frac{m_1+|m_1|}{2}}
\sqrt{(l+1/2)|m_2|} \, W(ab ed;\ cf),
\end{eqnarray}
where
\begin{center}
\begin{tabular}{lll}
$a = \frac{J+|m|-1}{4}$, \,\,
&
$b = \frac{J-|m|-2}{4}$, \,\,
&
$c = \frac{2l-1}{4}$,
\\[2mm]
$d = \frac{J-|m|}{4}$, \,\,
&
$e = \frac{J+|m|-1}{4}$, \,\,
&
$f = \frac{|m_2|-1}{2}$. \\
\end{tabular}
\end{center}
Let us mention an interesting fact that within the limit of a free motion
in formula (32) for the interbasis coefficients $W_{Nlm}^{n_3}(\nu)$
instead of the hypergeometric function ${_3F_2}$ of the unit argument
we have the function ${_4F_3}$; and instead of the Clebsch--Gordan
coefficients of the SU(2) group, the Racah coefficients for one fourth
values of the SU(1,1) group momentum. Analogous formulae arose in
calculating the coefficients of transition between different
hyperspherical systems of coordinates (in the formalism of "trees")
and have been analysed in \cite{KUZ}.
In our case, this fact allows an alternative
calculation of the integral
$A_{N|m|n_3}^{l}(\nu)$ in formula (28) at $\nu = 0$.

\section{Elliptic bases}

The oblate elliptic system of coordinates ( known as the
elliptic--cylindrical I) has the form

\begin{eqnarray*}
\begin{array}{ll}
  q_1 = R\sn(\mu,k)\dn(\nu,k')\cos\phi,  &
  q_2 = R\sn(\mu,k)\dn(\nu,k')\sin\phi, \\
  q_3 = R\cn(\mu,k)\cn(\nu,k'),  &
  q_0 = R\dn(\mu,k)\sn(\nu,k').
\end{array}
\end{eqnarray*}
$$
-K\leq \mu\leq K, -2K'\leq\nu\leq2K', 0\leq\phi<2\pi,
$$
where the elliptic Jacobi functions of the variables $\alpha$ and $\beta$
have the moduli $k$ and $k'$, respectively, $k^2 + k'^2 = 1$, and  $K$ and $K'$
are the complete elliptic integrals.

For the potential $V$ in the elliptic system of coordinates we have
\begin{equation}
\label{E1}
V(\mu, \nu) = \frac{1}{2} M{\omega}^2 R^2
\left[\frac{1}{\dn^2\mu \sn^2\nu}
- 1 \right].
\end{equation}
Choosing the wave function $\Psi$ in the form
\begin{equation}
\label{E2}
\Psi(\mu,\nu,\varphi) = {\psi}_1(\mu) {\psi}_2(\nu)
\frac{e^{im\varphi}}{\sqrt {2\pi}}, \qquad m \in\mbox{\rm\bf{Z}},
\end{equation}
after the separation of variables in the Schr\"odinger equation (\ref{E1})
we arrive at two ordinary differential equations
\begin{eqnarray}
\label{E3}
       \frac{d^2  {\psi}_1}{d\mu^2 } +
\frac{\cn\mu \dn\mu}{\sn\mu}\frac{d {\psi}_1}{d\mu}
- \Biggl[\left(\frac{2MER^2}{\hbar^2} +
\frac{M^2\omega^2R^4}{\hbar^2}\right)k^2 \sn^2\mu
&+& \frac{m^2}{\sn^2\mu}
- \frac{M^2\omega^2R^4}{\hbar^2}
\frac{k'^2}{\dn^2\mu}\Biggr] {\psi}_1
\nonumber\\[2mm]
&=& - \lambda_q (k;R) {\psi}_1,
\\ [2mm]
\label{E4}
       \frac{d^2  {\psi}_2}{d\nu^2} -
k'^2 \frac{\sn\nu \cn\nu}{\dn\nu}\frac{d  {\psi}_2}{d\nu}
+ \Biggl[\left(\frac{2MER^2}{\hbar^2} +
\frac{M^2\omega^2R^4}{\hbar^2}\right)\dn^2\nu
&+&\frac{k^2 m^2}{\dn^2\nu}
- \frac{M^2\omega^2R^4}{\hbar^2}
\frac{1}{\sn^2\nu}\Biggr] {\psi}_2
\nonumber\\[2mm]
&=& + \lambda_q (k;R) {\psi}_2,
\end{eqnarray}
where the quantum number $q$ enumerates the elliptic separation constant
$\lambda_q (k;R)$. Excluding from the equations (\ref{E3}) and (\ref{E4})
energy $E$, we come to the following operator
\begin{eqnarray}
\label{E6}
\Lambda
&=&
\frac{1}{k^2\sn^2\mu-\dn^2\nu}
\left[{\dn^2\nu} \frac{\partial^2}{\partial \mu^2}
+ k^2\sn^2\mu \frac{\partial^2}{\partial \nu^2}
+ \frac{\cn\mu \dn\mu}{\sn\mu} \dn^2\nu
\frac{\partial}{\partial \mu} -
k'^2k^2 \frac{\sn\nu \cn\nu}{\dn\nu}
\frac{\partial}{\partial \nu} \right]
\nonumber\\ [2mm]
&+&
\frac{M^2\omega^2 R^4}{\hbar^2}
\,\frac{\dn^2\nu + k^2\sn^2\mu -1}{\dn^2\mu \sn^2\nu}
- \frac{\dn^2\nu + k^2\sn^2\mu}{\dn^2\nu \sn^2\mu}
\,\frac{\partial^2}{\partial \varphi^2}
\nonumber\\[2mm]
&=& (1-k^2) L^2 - k^2 R^2 D_{33} + k^2 L_3^2
+ (k'^2) \frac{M^2 \omega^2 R^4}{\hbar^2} +
k^2 \frac{2MR^2}{\hbar^2} H,
\end{eqnarray}
whose eigenvalues are $\lambda_q(k;R)$, and eigenfunctions
are given by expression (\ref{E2}). Let us introduce a new operator
according to
\begin{eqnarray}
\label{E7}
\Im^{\mbox{\tiny{obl.}}}
&=& \frac{1}{1-k^2} \left\{
\Lambda +(1-k^2) \frac{M^2 \omega^2 R^4}{\hbar^2}
-k^2 \frac{2MR^2}{\hbar^2} H + k^2 L_3^2\right\}\nonumber
\\[2mm]
&=&L^2- a R^2 D_{33},
\end{eqnarray}
where $a = k^2/(1-k^2) \in [0,\infty)$. As is known \cite{GPKS},
a transition from the oblate to the prolate elliptic system of coordinates
on the sphere can be obtained under the transformation
$k\to i k/k',\,\, k'\to 1/k'$. In this case, the operator (\ref{E7}) for
the prolate system of coordinates has the form
\begin{equation}
\label{E8}
\Im^{\mbox{\tiny{prl.}}} = L^2 + k^2 R^2 D_{33}.
\end{equation}
The latter formula allows one to describe both the elliptic systems
of coordinates uniquely with the use of the operator
\begin{equation}
\label{E9}
\Im = L^2 - a R^2 D_{33},
\end{equation}
where $a\in [-1,\infty)$. For positive $a$ we have the oblate system;
and for $a\in [-1, 0]$, the prolate elliptic system of coordinates.

\subsection{ Expansion of elliptic bases over the hyperspherical
and cylindrical ones}

Thus, we have seen in the previous sections that all three oscillator
bases: hyperspherical, cylindrical and both the elliptic ones, are
eigenfunctions of three complete sets of operators
$\{H, L^2, L_z^2\}$, $\{H, D_{33}, L_z^2\}$ and
$\{H, \Im, L_z^2\}$  so that
\begin{eqnarray}
\label{R1}
L^2{\Psi}_{Nlm}(\chi,\vartheta,\varphi)
&=& l(l+1){\Psi}_{Nlm}(\chi,\vartheta,\varphi),
\\[2mm]
\label{R2}
D_{33}{\Psi}_{N n_3 m}(\varphi_1,\alpha,\varphi_2) &=& (n_3+\nu+1)^2
{\Psi}_{N n_3 m}(\varphi_1,\alpha,\varphi_2),
\\[2mm]
\label{R3}
\Im {\Psi}_{nqm}(\mu,\nu,\varphi) &=&
{\lambda}_q(a;R) {\Psi}_{nqm}(\mu,\nu,\varphi).
\end{eqnarray}
The operator equations (\ref{R1})-(\ref{R3}) allow us to construct
elliptic bases of the isotropic oscillator on the sphere as a superposition
over the hyperspherical and cylindrical bases.

Now, let us write the sought expansions:
\begin{eqnarray}
\label{R5}
{\Psi}_{Nqm}(\mu,\nu,\varphi) &=&
\sum_{l=|m|,|m|+1}^{N}{T}_{Nqm}^{l}(a;R)
{\Psi}_{N l m}(\chi,\vartheta,\varphi),
\\[2mm]
\label{R4}
{\Psi}_{Nqm}(\mu,\nu,\varphi) &=&
\sum_{n_3=0,1}^{N-|m|}{U}_{Nqm}^{n_3}(a;R)
{\Psi}_{N n_3 m}(\varphi_1, \alpha, \varphi_2).
\end{eqnarray}
Consider the expansion (\ref{R5}). Substituting (\ref{R5}) into
the operator equation (\ref{R2}), we find
\begin{eqnarray}
\label{R6}
\frac{1}{a R^2}\left\{l(l+1)-\lambda_q(a;R)\right\}
T_{Nqm}^l= \sum_{l'=0}^N T_{Nqm}^{l'}(D_{33})_{l l'},
\end{eqnarray}
where
\begin{equation}
\label{R7}
(D_{33})_{l l'} = \int {\Psi}_{N l m}^{*}(D_{33}){\Psi}_{N l' m}
d \Omega.
\end{equation}
To calculate the integral (\ref{R7}) we use the expansion of the
spherical basis over the cylindrical one and equation (42)
for the eigenfunctions of the operator $D_{33}$. As a result, we
come to the following expression for $(D_{33})_{l l'}$:
\begin{eqnarray}
\label{R9}
(D_{33})_{l l'}
=\sum_{n_3}^{N-|m|}{W}_{Nlm}^{n_3}{W}_{Nl'm}^{n_3} (n_3+\nu+1)^2,
\end{eqnarray}
Then, using the three-term recurrence relations for the Racah
coefficients \cite{V}
\begin{eqnarray}
\label{R10}
c B_c \cases{a, b, c+1 \cr d, l, f \cr} \bigg\}
+(c+1) B_{c-1}\cases{a, b, c-1 \cr d, l, f\cr} \bigg\}
+(2 c+1) A_c \cases{a, b, c \cr d, l, f \cr} \bigg\} = 0,
\end{eqnarray}
where
\begin{eqnarray}
\label{R11}
B_c&=&\sqrt{(a+b+c+2)(-a+b+c+1)(a-b+c+1)(a+b-c)}
\\[2mm]
&\times&\sqrt{(d-l+c+1)(d+l-c)(d+l+c+2)(-d+l+c+1)},\nonumber
\\[2mm]
\label{R12}
A_c&=&[a(a+1)-b(b+1)][d(d+1)-l(l+1)]+c(c+1)[a(a+1)
\\[2mm]
&+& b(b+1)+ d(d+1)+l(l+1)-c(c+1)]
-2c(c+1)f(f+1)\nonumber
\end{eqnarray}
and the orthogonality property
\begin{eqnarray}
\label{R19}
\sum_{n_3=0,1}^{N-|m|}{W}_{Nlm}^{n_3}{W}_{Nl'm}^{n_3} = \delta_{l l'},
\end{eqnarray}
we have
\begin{eqnarray}
\label{R20}
(D_{33})_{l l'} =
-\frac{16 B_{l-2}}{(2l-1)(2l+1)} \delta_{l-2, l'}\nonumber
+C_l \delta_{l l'}
-\frac{16 B_{l}}{(2l+1)(2l+3)} \delta_{l+2, l'}
\end{eqnarray}
where
\begin{eqnarray}
\label{R15}
B_l&=&\frac{1}{16}\sqrt{(l-|m|+1)(l-|m|+2)(l+|m|+1)
(l+|m|+2)(N+l+3)(N-l)}\nonumber
\\[2mm]
&\times&\sqrt{(N+l+2\nu+4)(N-l+2\nu+1)},
\\[2mm]
\label{R16}
C_l&=&\frac{1}{8}\bigg\{
4 (N+1) (N+3)+2 (2 |m|^2-1)+ 4\nu(2 N+2 \nu+5)-(2l-1)(2l+3)\nonumber
\\[2mm]
&-&\frac{(4|m|^2-1)(2N+3)(2 N+5+\nu)}{(2l-1)(2l+3)}
\bigg\}
\end{eqnarray}
Substituting the matrix element (\ref{R20}) into (\ref{R6}),
we finally arrive at the three-term recurrence relation
\begin{eqnarray}
\label{R21}
\frac{16}{(2 l-1)(2 l +1)}B_{l-2} T_{Nqm}^{l-2}(a;R)
+\left\{\frac{1}{a R^2}\left[l(l+1)-\lambda_q(a;R)\right]-C_l\right\}
T_{Nqm}^l(a;R)
\nonumber
\\[2mm]
+\frac{16}{(2 l+1)(2 l +3)} B_lT_{Nqm}^{l+2}(a;R) = 0,
\\[2mm]
T_{Nqm}^{-1}(a;R) = T_{Nqm}^{-2}(a;R) = 0.
\nonumber
\end{eqnarray}
for the expansion coefficients $T_{Nqm}^{l}(a;R)$. The recurrence
relation (\ref{R21}) is a system of homogeneous equations which has
to be solved with the normalization condition
\begin{eqnarray*}
\sum_{l=|m|, |m|+1}^{N}|T_{Nqm}^l(a;R)|^2 = 1.
\end{eqnarray*}
Eigenvalues of the elliptic separation constant $\lambda_q(a;R)$
are calculated from the condition for the determinant
of the system of homogeneous equations to be equal to zero (\ref{R21}).

Consider now the expansion (\ref{R4}) of the elliptic basis over the
cylindrical one. In a similar way, like in the calculation of the expansion
coefficients (\ref{R5}), we get
\begin{eqnarray}
\label{R22}
\left\{\lambda_q(k;R)+a R^2(n_3+\nu+1)^2\right\}U_{N q m}^{n_3}(k;R) =
\sum_{n_3'=0,1}^{N-|m|} U_{N q m}^{n_3'}(k;R)
(L^2)_{n_3 n_3'},
\end{eqnarray}
where
\begin{equation}
\label{R23}
(L^2)_{n_3 n_3'} =
\int {\Psi}_{N n_3 m}^{*} L^2
{\Psi}_{N n_3' m}d \Omega.
\end{equation}
The integral in (\ref{R23}) can be calculated if one uses the expansion
of the cylindrical basis over the spherical one and then the symmetry
property of the Racah coefficients
\begin{eqnarray}
\label{R26}
\cases{a, b, c \cr d, l, f \cr} \bigg\}
= \cases{a, l, f \cr d, b, c\cr} \bigg\}
\end{eqnarray}
and the three-term recurrence relation (\ref{R10}).
As a result of simple calculations we obtain the expression
($U_{N q m}^{n_3}\equiv U_{n_3}$)
\begin{eqnarray}
\label{R28}
\tilde B_{n_3} U_{n_3+2}+\left\{\tilde C_{n_3}-
\lambda_q(k;R)-a R^2(n_3+\nu+1)^2
\right\} U_{n_3}+\tilde B_{n_3-2}U_{n_3-2}=0,
\end{eqnarray}
where
\begin{eqnarray*}
\tilde B_{n_3}&=&\frac{1}{4}\sqrt{
\frac{(n_3+2 \nu+2)(n_3+2)(n_3+1)(n_3+2 \nu+3)(N+|m|+n_3+2\nu+4)}
{(n_3+\nu+1)(n_3+\nu+2)^2(n_3+\nu+3)}}
\\[2mm]
&\times&\sqrt{(N+|m|-n_3)(N-|m|-n_3)(N-|m|+n_3+2\nu+4)},
\\[2mm]
\tilde C_{n_3}&=&\frac{1}{2}\bigg\{(N+2)^2+\nu(2 N +2\nu+5)+(|m|^2-2)
-(n_3+\nu)(n_3+\nu+2)
\\[2mm]
&-&\frac{\nu(\nu+1)(N+|m|+\nu+2)(N-|m|+\nu)}{(n_3+\nu)(n_3+\nu+2)}
\bigg\}.
\end{eqnarray*}
As in the previous case, a homogeneous system of equations should be
solved together with the normalization condition
\begin{eqnarray*}
\sum_{n_3=0,1}^{N-|m|} |U_{Nqm}^{n_3}(a;R)|^2 = 1,
\end{eqnarray*}
and the separation constant can again be determined from the condition of
equality to zero of the corresponding determinant of the homogeneous system
of equations (\ref{R28}).

In conclusion, we would like to mention that in the limit of the free
motion $\nu\to 0$, the three-term recurrence relations (55) and (59)
transform into those for the expansion coefficients of the elliptic
basis over cylindrical and spherical ones, which
have been obtained in the work [27].

\section{Conclusion}

In the present paper, a first step has been made to completely study
the Schr\"odinger equation and to calculate the coefficients of various
interbasis expansions for the potential of the isotropic oscillator
on the three-dimensional sphere in different orthogonal systems of coordinates.
We have calculated interbasis expansions for the "sphere--cylinder"
transition and also constructed solutions of the Schr\"odinger equation
in both the elliptic systems of coordinates as expansion over the
spherical and cylindrical bases. In contrast with the "sphere--cylinder"
transition, in which the transformation coefficients are expressed through
the generalised hypergeometric functions $_4F_3$ of the unit argument
or through the Racah coefficients, extended over their indices into the
region of arbitrary real values, the transition coefficients for the elliptic
bases are defined by the three-term recurrence relations and cannot be
written down explicitly.

In this paper we have considered the sphero-conic and ellipsoidal bases
of the isotropic oscillator. In separating variables in the Schr\"odinger
equation for the sphero-conic system of coordinates we arrive at the
quasiradial equation (13) considered in Sect. 3 and two standard Lame
equations that are derived in separating variables for the Helmholtz
equation on the two-dimensional sphere. As a result, the solution of the
Schr\"odinger equation is
\begin{eqnarray*}
\Psi(\chi, \alpha, \beta; R)  =
Z^{\nu}_{Nl}(\chi) \, \Lambda_{l\lambda}(\alpha)
\Lambda_{l\lambda}(\beta),
\end{eqnarray*}
where the function $Z^{\nu}_{Nl}(\chi)$ is determined by expression (16), and
the explicit form of the Lame polynomials $\Lambda_{l\lambda}(\alpha)$ can
be found in \cite{BE,PW}.

For the ellipsoidal basis of the isotropic oscillator, after separation of
variables in the Schr\"odinger equation we obtain three identical
equations containing two ellipsoidal separation constants
$\lambda_1,\lambda_2$. This means that in contrast with the cases considered
we deal with the two-parametric spectral problem. Application of the method
of constructing solutions of the Schr\"odinger equation as expansion over
simpler bases leads to two many-termed recurrence relations determined by
the cubic matrix. Obviously, the simplest way of constructing an ellipsoidal
basis consists in applying the Niven method \cite{WW} allowing one to
write down a solution to the Schr\"odinger equation in terms of zeroes of the
wave function and reduce the problem to a system of relevant nonlinear
equations. This kind of investigation is beyond the scope of the present
paper and will be carried out elsewhere.

In conclusion, we would like to thank L.G. Mardoyan and V.M. Ter-Antonyan
for useful interesting discussions.


\begin{thebibliography}{99}

\bibitem{SCH}
Schr\"odinger E., {\it Proc. of the Royal Irish.Acad.} 1940, {\bf A46}, p. 9;
p. 183; 1941, {\bf A47}, p. 53.
\bibitem{IH}
Infeld L. and Schild A.,
{\it Phys.Rev.} 1945, {\bf 67}, p. 121.
\bibitem{STIV}
Stevenson A.F.,
{\it Phys.Rev.} 1941, {\bf 59}, p. 842.
\bibitem{BB}
Bessis N. and Bessis G., {\it J.Phys.} 1979, {\bf A12}, p. 1991.
\bibitem{IZ}
Izmest'ev A.A., {\it Yad. Fiz.} 1991, {\bf 52}, p. 1697; 1991,
{\bf 53}, p. 1402 (in Russian).
\bibitem{BOT}
Bogush A.A., Otchik V.S., {\it J.Phys.} 1997, {\bf A30}, p. 559.
\bibitem{NI}
Nishino Y., {\it Math. Japon.} 1972, {\bf 17}, p. 59.
\bibitem{HI}
Higgs P.W., {\it J.Phys.} 1979, {\bf A12}, p. 309.
\bibitem{LE}
Leemon H.I., {\it J.Phys.} 1979, {\bf A12}, p. 489.
\bibitem{KO}
Kurochkin Y.A., Otchik V.S., {\it DAN BSSR} 1979, {\bf XXIII}, p. 987
(in Russian).
\bibitem{BKO}
Bogush A.A., Kurochkin Y.A., Otchik V.S., {\it DAN BSSR} 1980,
{\bf XXIV}, p. 19
(in Russian).
\bibitem{BOR}
Bogush A.A., Otchik V.S., Red'kov V.M., {\it Vestnik AN BSSR} 1983, {\bf 3},
p. 56 (in Russian).
\bibitem{OR}
Bogush A.A., Red'kov V.M.,
{\it Quantum -- Mechanical Kepler problem in constant curvature
spaces},
Preprint 298, {\it IN AN BSSR} 1983
(in Russian).
\bibitem{GZL1}
Granovsky Ya.A., Zedanov A.S., Luzenko I.M.,
{\it Teor.Mat.Fiz.} 1992, {\bf 91}, p. 207
(in Russian).
\bibitem{GZL2}
Granovsky Ya.A., Zedanov A.S., Luzenko I.M.,
{\it Teor.Mat.Fiz.} 1992, {\bf 91}, p. 396
(in Russian).
\bibitem{BIJ1}
Barut A.O., Inomata A. and Junker G.,
{\it J.Phys.} 1987, {\bf A20}, p. 6271.
\bibitem{BIJ2}
Barut A.O., Inomata A. and Junker G.,
{\it J.Phys.} 1990, {\bf A23}, p. 1179.
\bibitem{GR}
Grosche C., {\it Ann. Phys. (N.Y.)} 1990, {\bf 204}, p. 208.
\bibitem{GRO1}
Grosche C., Pogosyan G.S. and Sissakian A.N.,
{\it Fortschritte der Physik}, 1995, {\bf 43}, p. 523.
\bibitem{GRO2}
Grosche C., Pogosyan G.S. and Sissakian A.N.,
{\it Phys.Part.Nucl.} 1996,
{\bf 27}(3), p. 244.
\bibitem{GRO3}
Grosche C., Pogosyan G.S. and Sissakian A.N.,
{\it Phys.Part.Nucl.} 1997,
{\bf 28}(5).
\bibitem{OL}
Olevskii, {\it Mat. Sb.} 1950, {\bf 27}, p. 379
(in Russian).
\bibitem{DEM}
Demkov Yu.N., {\it JETP} 1965, {\bf 36}, p. 88
(in Russian).
\bibitem{KAD}
Kadyshevsky V.G., {\it Dok.Akad.Nauk. SSSR} 1962, {\bf 147},
p. 588 (in Russian).
\bibitem{POG1}
Pogosyan G.S. and Sissakian A.N., {\it Turkish J.Phys.} 1997, {\bf 21}, p. 515.
\bibitem{IPSW}
Izmest'ev A.A., Pogosyan G.S., Sissakian A.N. and Winternitz P.,
{\it J.Phys.} 1996, {\bf A29}, p. 5949.
\bibitem{GPKS}
Grosche C., Karayan Kh.H., Pogosyan G.S. and Sissakian A.N.,
{\it J.Phys.} 1997, {\bf A30}, p. 1629.
\bibitem{BE}
Bateman H., Erd\'elyi A., {\it Higher transcendental functions},
New York--Toronto--London, MC Graw--Hill Book Company, INC, 1953.
\bibitem{BA}
Bailey W.N., {\it Generalized hypergeometric series}, Cambridge Tracts, No 32,
Cambridge, 1935.
\bibitem{P}
Prudnikov A.P., Brichkov Yu.A., Marichev O.I.,
{\it Integrals and series. Special functions.}, Nauka, Moskva, 1986
(in Russian).
\bibitem{V}
Varshalovich D.A., Moskalev A.N. and Khersonskii V.K.,
{\it Quantum theory of angular momentum}, World Scientific, Singapore, 1988.
\bibitem{KUZ}
Kuznetsov G.I., Smorodinsky Ya.A., {\it Yad. Fiz.} 1977, {\bf 25}, p. 447.
(in Russian).
\bibitem{PT}
Pogosyan G.S., Ter--Antonyan V.M.,
{\it Conversion Factors for Cartesian, Polar and Cylindrical
wave functions of Isotropic Oscillator.} Communication JINR 1978,
P2-11962, Dubna
(in Russian).
\bibitem{S}
Szego G., {\it Orthogonal Polynomials}, Amer.Math.Society, New York,
1959.
\bibitem{FL}
Fl\"ugge S., {\it Practical Quantum Mechanics}, Springer--Verlag,
Berlin--Heidelberg--New York, 1971.
\bibitem{PW}  Patera J. and Winternitz P.,
{\it J.Math.Phys.} 1973, {\bf 14}, p. 1130.
\bibitem{WW}
Whittaker E.T., Watson G.N.,
{\it A course of modern analysis}, Cambridge University Press 1927, {\bf 2}.
\end{thebibliography}
\end{document}